\documentclass[twocolumn,showpacs,preprintnumbers,amsmath,amssymb]{revtex4}
\usepackage{graphicx}
\usepackage{graphicx}
\usepackage{dcolumn}
\usepackage{bm}

\begin{document}

\preprint{PUPT-2239, LMU-ASC 42/07}

\title{
Sonic booms and diffusion wakes generated by a heavy quark in thermal AdS/CFT
}

\author{Steven S. Gubser}\email{ssgubser@Princeton.EDU}
\author{Silviu S. Pufu}\email{spufu@Princeton.EDU}
\affiliation{Joseph Henry Laboratories, Princeton University, Princeton, NJ 08544}
\author{Amos Yarom}\email{yarom@theorie.physik.uni-muenchen.de}
\affiliation{Ludwig-Maximilians-Universit\"at, Department f\"ur Physik, Theresienstrasse 37, 80333 M\"unchen, Germany}

\date{\today}

\begin{abstract}
We evaluate the Poynting vector generated by a heavy quark moving through a thermal state of ${\cal N}=4$ gauge theory using AdS/CFT\@.  A significant diffusion wake is observed as well as a Mach cone.  We discuss the ratio of the energy going into sound modes to the energy coming in from the wake.
\end{abstract}

\pacs{11.25.Tq, 25.75.-q}

\maketitle

\section{Introduction}

The phenomenon of jet-splitting observed at the Relativistic Heavy Ion Collider (RHIC) \cite{Adler:2005ee,Adams:2005ph} motivates a study of the stress tensor generated by a quark moving through a thermal state of ${\cal N}=4$ super-Yang-Mills theory (SYM)\@.  Although SYM is significantly different from quantum chromodynamics (QCD), it is widely hoped that comparisons between the two will lead to insights not readily extracted from conventional techniques such as perturbation theory and lattice simulations.  The advantage of studying SYM is that powerful string theory methods exist to calculate gauge-singlet observables, such as the stress tensor, at strong coupling, starting from gravitational calculations in five-dimensional anti-de Sitter space (AdS${}_5$).  These methods rely on the fact that SYM is a conformal field theory (CFT).

The basic string theory setup is well reviewed in earlier works \cite{Friess:2006aw,Friess:2006fk,Yarom:2007ni,Gubser:2007nd,Yarom:2007ap,Gubser:2007xz,Chesler:2007an}.  Briefly, an infinitely massive, fundamentally charged quark that is dragged at constant velocity through an infinite, static, thermal plasma of SYM can be thought of as having a string trailing behind it down into AdS${}_5$-Schwarzschild, generating a drag force as computed in \cite{Herzog:2006gh,Gubser:2006bz} (see also \cite{Liu:2006ug,Casalderrey-Solana:2006rq}).  This string perturbs the metric $g_{\mu\nu}$ of AdS${}_5$-Schwarzschild, and those perturbations can be translated into the expectation value $\langle T^{mn} \rangle$ of the stress tensor in the boundary gauge theory.  Although it would perhaps be preferable to study an energetic gluon propagating through the thermal plasma, the infinitely heavy quark case is a textbook exercise in AdS/CFT because the quark has no dynamics apart from its constant-velocity motion: thus its trajectory can be regarded as part of the boundary conditions specified in the gravitational calculation.  Infalling boundary conditions are specified at the horizon of AdS${}_5$-Schwarzschild, corresponding to causal physics in the boundary gauge theory.

In the first work on jet-splitting in AdS/CFT \cite{Friess:2006fk}, Fourier coefficients $\langle T^{mn}(\vec{k}) \rangle$ were computed numerically, and high-$k$ and low-$k$ asymptotics were given.  Subsequent work \cite{Yarom:2007ap,Gubser:2007nd,Yarom:2007ni,Gubser:2007xz} has considerably refined the high-$k$ estimates.  The position-space stress tensor can in principle be recovered from its Fourier coefficients using
 \begin{equation}
  \langle T^{mn}(t,\vec{x}) \rangle =
    \int {d^3 k \over (2\pi)^3} e^{i k_1 (x^1 - vt) + i k_2 x^2 + i k_3 x^3}
     \langle T^{mn}(\vec{k}) \rangle \,. \label{ThreeFourier}
 \end{equation}
Here and below, $v$ is the quark's velocity directed along the positive $x^1$ direction.
Evidence of a sonic boom was already exhibited in \cite{Friess:2006fk}, and in \cite{Gubser:2007xz,Chesler:2007an} it was made more concrete by evaluating the full position-space energy density $\langle T^{00}(t,\vec{x}) \rangle$.  A match to linearized hydrodynamics with $\eta/s = 1/4\pi$ was remarked upon in \cite{Gubser:2007xz,Chesler:2007an}.  What AdS/CFT provides beyond hydrodynamics is an all-scales description of dissipation, from the near-field of the quark all the way out to the linearized hydro regime.

A crucial question for application to RHIC phenomenology is the relative strength of the sonic boom and the diffusion wake.  In the lab frame, the diffusion wake is a structure primarily visible in the Poynting vector, and it is concentrated within a parabolic region behind the quark.  It is closely related to the well-known laminar wake of non-relativistic hydrodynamics.  The forward flow in the diffusion wake should lead to enhanced particle production approximately collinear with the hard parton, as compared to vacuum fragmentation.  The sonic boom, in contrast, describes high-angle emission.  Intriguingly, data from the PHENIX experiment \cite{Adler:2005ee} favors suppression of the diffusion wake to such an extent that earlier authors \cite{Casalderrey-Solana:2006sq} have suggested turning it off altogether in a linearized hydro analysis.  Other authors have proposed a model \cite{Renk:2006mv} in which it is assumed that $75\%$ percent of energy lost goes into a sonic boom.  Data from STAR does not show as clear a minimum in associated particle production in the direction of the away-side parton \cite{Adams:2005ph}.  In the string theory calculation, the relative strength of the sonic boom and the diffusion wake is fixed.  Because other first-principles estimates of the relative strength of these two effects are lacking, it is clearly of interest to work out the string theory predictions in some detail.  All the pieces are already in place except for sufficiently precise high-$k$ estimates of $\langle T^{0i}(\vec{k}) \rangle$ and a robust numerical scheme for performing the Fourier transforms.  In this letter, we fill in these last gaps and extract explicit results for the Poynting vector.  We also discuss one way of quantifying the relative strength of sonic and diffusive modes which suggests that, in the rest frame of the plasma, the sonic boom dissipates $1+v^2$ times as much energy as the diffusion wake feeds in.

\section{Definitions}

The only scale in the problem is the temperature, so we will consistently work with dimensionless quantities
 \begin{eqnarray}
  \vec{X} &\equiv& \pi T \vec{x} \\
  \vec{K} &\equiv& \vec{k} / \pi T \,.
 \end{eqnarray}
Instead of computing the expectation value of the complete stress tensor, we focus on a rescaled, subtracted quantity
 \begin{eqnarray}
  &{\cal T}^{mn}(\vec{X}) \equiv
    {\sqrt{1-v^2} \over (\pi T)^4 \sqrt{g_{YM}^2 N}}
      \Big( \langle T^{mn}(0,\vec{x}) \rangle -
       \langle T^{mn} \rangle_{\rm bath} \Big) \label{CurlyT} \\
  &\langle T^{mn} \rangle_{\rm bath} \equiv
    {\pi^2 \over 8} (N^2-1) T^4 {\rm diag}\{3,1,1,1\} \,.
 \end{eqnarray}
On the right hand side of (\ref{CurlyT}) we have set $t=0$, which leads to no loss of information because the computations are all done in a steady state approximation.  The powers of $\pi T$ render ${\cal T}^{mn}$ dimensionless.  The factor of $1/\sqrt{g_{YM}^2 N}$ cancels out an overall scaling with $\sqrt{g_{YM}^2 N}$ of all disturbances due to the trailing string.  This scaling arises because it's how the string tension depends on $g_{YM}$ and $N$.  The quantity $\langle T^{mn} \rangle_{\rm bath}$ is just the stress tensor of the SYM plasma in the absence of the quark.

Our focus is on the first row of ${\cal T}^{mn}$, namely
 \begin{equation}
  {\cal T}^{0m} \equiv \begin{pmatrix} {\cal E} & {\cal S}_1 &
    {\cal S}_2 & {\cal S}_3 \end{pmatrix} \,,
 \end{equation}
where ${\cal E}$ and $\vec{\cal S}$ are the rescaled, bath-subtracted energy density and Poynting vector.  (We work in signature $-$+++, so there is no distinction between ${\cal S}^i$ and ${\cal S}_i$.)  We further define
 \begin{eqnarray}
  E &\equiv& {\cal E} - {\cal E}_{\rm Coulomb} \\
  \vec{S} &\equiv& \vec{\cal S} - \vec{\cal S}_{\rm Coulomb} \,,
 \end{eqnarray}
where ${\cal E}_{\rm Coulomb}$ and ${\cal S}_{\rm Coulomb}$ characterize the Coulombic near-field of the quark.  Thus $E$ and $\vec{S}$ exclude not only the contribution of the bath in the absence of the quark, but also the contribution of the quark in the absence of the bath.  The quantities which we eventually plot are $E$, $S_1$, and $S_\perp$ as functions of $X_1$ and $X_\perp$.  Here
 \begin{equation}
  X_\perp \equiv \sqrt{X_2^2 + X_3^2} \,,
 \end{equation}
and similarly for $K_\perp$ and $S_\perp$.

\section{Asymptotics and subtractions}

The key to a robust numerical evaluation of stress-tensor components in position space is to make a split
 \begin{eqnarray}
  {\cal E} &=& {\cal E}^{\rm UV} + {\cal E}^{\rm IR} +
    {\cal E}^{\rm res}  \\
  \vec{\cal S} &=& \vec{\cal S}^{\rm UV} + \vec{\cal S}^{\rm IR} +
    \vec{\cal S}^{\rm res} \,,
 \end{eqnarray}
where ${\cal E}^{\rm UV}$ and $\vec{\cal S}^{\rm UV}$ are analytical approximations to ${\cal E}$ and $\vec{\cal S}$ for large $K$ which are valid to order $K^{-3}$, ${\cal E}^{\rm IR}$ and $\vec{\cal S}^{\rm IR}$ are analytical approximations for small $K$ valid to order $K^0$, and the remainders ${\cal E}^{\rm res}$ and $\vec{\cal S}^{\rm res}$ are uniformly bounded and integrable, so that they can be passed through a fast Fourier transform (FFT) with controllable errors.  The full expressions for the analytical approximations are too long to be presented in full, but the leading terms are
 \begin{align}
  {\cal E}^{\rm UV} &= -{2+v^2 \over 24} \sqrt{K_1^2 (1-v^2) +
     K_\perp^2 + \mu_{\rm UV}^2} \nonumber \\
    &\ {} +
     {2 v^2 K_1^2 (1-v^2) + (2+v^2) \mu_{\rm UV}^2 \over
     48 \sqrt{K_1^2 (1-v^2) + K_\perp^2 + \mu_{\rm UV}^2}} +
     {\cal O}(K^{-1}) \label{EUV}\\
  {\cal S}_1^{\rm UV} &= -{v \over 8} \sqrt{K_1^2 (1-v^2) +
     K_\perp^2 + \mu_{\rm UV}^2} \nonumber \\
   &\ {} +
     {2 v K_1^2 (1-v^2) + 3 v \mu_{\rm UV}^2 \over
     48 \sqrt{K_1^2 (1-v^2) + K_\perp^2 + \mu_{\rm UV}^2}} +
     {\cal O}(K^{-1}) \label{S1UV} \\
  {\cal S}_2^{\rm UV} &= {v K_1 K_2 (1-v^2) \over
     24 \sqrt{K_1^2 (1-v^2) + K_\perp^2 + \mu_{\rm UV}^2}} +
     {\cal O}(K^{-1}) \label{SpUV}\\
  {\cal E}^{\rm IR} &=
     -{1\over 2\pi} {3 i v K_1 (1+v^2) - 3 v^2 K_1^2 \over K^2 -
     3 v^2 K_1^2 - i v K^2 K_1} \nonumber \\
   &\ {}
     +{1\over 2\pi} {3 i v K_1 (1+v^2) - 3 v^2 K_1^2 \over K^2 -
     3 v^2 K_1^2 - i v K^2 K_1 + \mu_{\rm IR}^2} \label{EIR} \\
  {\cal S}_1^{\rm IR} &=
     - {1\over 2\pi} {i (1+v^2) K_1 + v K^2 - 2 v^3 K_1^2 \over
    K^2 - 3 v^2 K_1^2 - i v K^2 K_1} \nonumber \\
   &\ {} + {2 v\over \pi} {1 + i K_1/4 v \over K^2 - 4 i v K_1}
   + \hbox{(regulators)} \label{S1IR} \\
  {\cal S}_2^{\rm IR} &=
    -{1\over 2 \pi} {i (1+v^2) K_2 + v^3 K_1 K_2 \over K^2 -
    3 v^2 K_1^2 - i v K^2 K_1} \nonumber \\
   &\ {} +
    {1\over 2\pi} {i K_2\over K^2 - 4 i v K_1} + \hbox{(regulators)}
     \label{SpIR}\,.
 \end{align}
Expressions for ${\cal S}_3^{\rm UV}$ and ${\cal S}_3^{\rm IR}$ may be deduced by rotational invariance around the $K_1$ axis.  The omitted terms in (\ref{S1IR}) and~(\ref{SpIR}) denoted ``$\hbox{(regulators)}$'' are analogous to the second term in (\ref{EIR}).

The strategy for obtaining UV subtractions was explained in \cite{Gubser:2007xz}: we solve for the metric perturbations at large $K$ using an iterative Green's function approximation, as first introduced in \cite{Yarom:2007ap}, then ``soften'' the resulting power series through the introduction of a dimensionless parameter $\mu_{\rm UV}$ as in (\ref{EUV})--(\ref{SpUV}).  To obtain the IR subtractions, we draw upon results of \cite{Friess:2006fk,Gubser:2007xz} to find series expansions at small $K$ for components of ${\cal T}^{mn}$, then partially resum the series, as in (\ref{EIR})--(\ref{SpIR}), to eliminate divergences on the Mach cone and the $X_1$ axis, then add Pauli-Villars-style regulators involving another dimensionless parameter $\mu_{\rm IR}$.

The terms in (\ref{EIR})--(\ref{SpIR}) with cubic denominators do not appear to admit analytical Fourier transforms; however they can easily be Fourier transformed in the $K_\perp$ directions, leaving a final one-dimensional FFT to be performed numerically.  Some of the subleading terms in our subtractions do not by themselves satisfy appropriate conservation properties, so the conservation of the full stress tensor has to be restored by combining numerical and analytical results.

\section{Conservation}

The stress tensor must be conserved except at the location of the quark, and the failure of conservation there matches the drag force on the quark \cite{Friess:2006fk}.  In terms of our rescaled variables, and in the rest frame of the plasma,
 \begin{eqnarray}
  {\partial \over \partial X^m} {\cal T}^{mn} &=& {\cal J}^n =
    -{\cal F}_{\rm drag}^n \delta(X^1-vX^0) \\
    \label{dTJ}
  {\cal F}_{\rm drag}^m &\equiv&
  -{1 \over 2\pi} \begin{pmatrix} v^2 & v & 0 & 0 \end{pmatrix} \,.
 \end{eqnarray}
After using the co-moving ansatz, (\ref{dTJ}) becomes
 \begin{eqnarray}
  &i K_m {\cal T}^{mn}(\vec{K}) = -{\cal F}_{\rm drag}^n  \\
  &K_m = \begin{pmatrix} -v K_1 & K_1 & K_2 & K_3 \end{pmatrix} \,.
 \end{eqnarray}
Now consider a split of the IR asymptotic values of $\mathcal{T}^{mn}$
 \begin{equation}
  {\cal T}_{\rm IR}^{0m} = {\cal T}^{0m}_{\rm sound} +
    {\cal T}^{0m}_{\rm diffuse}  \label{TSplit}
 \end{equation}
where we define ${\cal T}^{0m}_{\rm sound}$ as all the terms in (\ref{EIR})--(\ref{SpIR}) with cubic polynomials in the denominator; ${\cal T}^{0m}_{\rm diffuse}$ by definition comprises all the remaining terms.  With this split,
 \begin{eqnarray}
  \lim_{K \to 0} i K_m {\cal T}_{\rm sound}^{0m} &=&
    {1 + v^2 \over 2\pi} =
    -\left( 1 + {1 \over v^2} \right) {\cal F}_{\rm drag}^0
    \label{SoundForce}\\
  \lim_{K \to 0} i K_m {\cal T}_{\rm diffuse}^{0m} &=&
    -{1 \over 2\pi} =
    {1 \over v^2} {\cal F}_{\rm drag}^0 \,. \label{DiffuseForce}
 \end{eqnarray}
The $m=0$ term in the left hand side of (\ref{SoundForce}) is, formally, the time derivative of the total energy in sound modes.  Adding in the $m=1,2,3$ terms accounts for energy flow into sound modes at large $X$, and the sum over all $m$ is the total rate of energy loss by the quark into sound modes.  Likewise, (\ref{DiffuseForce}) is the energy loss into diffusion modes.  But the relative sign between (\ref{SoundForce}) and~(\ref{DiffuseForce}) shows that while sound modes carry energy away from the quark, the diffusion wake actually feeds energy in toward the quark.

Energy loss to sound modes, as measured by (\ref{SoundForce}), is greater by a factor of $1+1/v^2$ than the total energy loss, and $1+v^2$ times the energy fed in by the diffusion wake \footnote{Another way of splitting ${\cal T}^{0m}$ into sound and diffusion modes is to associate ${\cal E}$ and the longitudinal component of the Poynting vector $\vec{\cal S} \cdot \hat{K}$ with the sound mode, and to associate the transverse part of $\vec{\cal S}$ (orthogonal to $\hat{K}$) with the diffusion mode.  This has the advantage that it decouples the equations of linearized hydrodynamics. This decomposition nearly agrees with the split discussed in (\ref{TSplit}), but additional non-local terms are added to the longitudinal modes and subtracted from the transverse modes which lead to non-zero but canceling values of $\vec{\cal S}^{\rm longitudinal}$ and $\vec{\cal S}^{\rm transverse}$ far ahead of the quark.  If this split is followed, then formally one finds zero net energy going into the transverse mode.  We thank D.~Teaney for discussions related to these points.}.  These ratios should be regarded with some caution because the argument leading to them is essentially algebraic; however, integrating the energy flux across the diffusion wake confirms (\ref{DiffuseForce}).

Instead of calculating in the rest frame of the plasma, as we have done, one may instead perform computations analogous to (\ref{SoundForce}) and (\ref{DiffuseForce}) in the rest frame of the quark.  Then, using also the IR asymptotics of $\mathcal{T}_{11}$ obtained in \cite{Friess:2006fk}, one can show that the energy lost by the quark into sound modes precisely cancels the energy fed in through diffusion modes. Some intuition regarding this result may be gained by noting that in the quark rest frame, the zero component of the drag force vanishes.

\section{Numerical results}

\begin{figure*}
  \hskip-0.5in\includegraphics[width=7.5in]{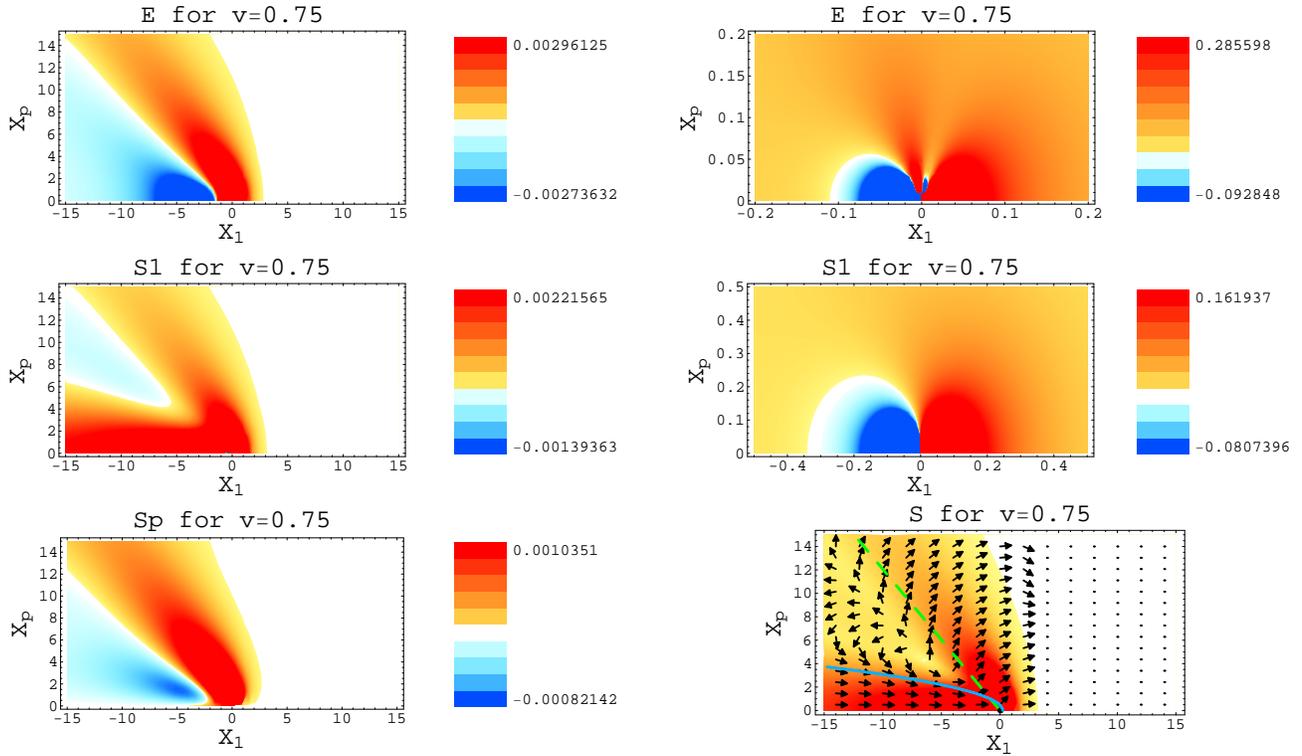}
 \caption{\label{BIGFIG}Energy density and components of the Poynting vector.  Sp means $S_\perp = \sqrt{S_2^2+S_3^2}$.  $|X|=1$ corresponds to $0.2\,{\rm fm}$ from the quark if $T \approx 318\,{\rm MeV}$.  Red means positive, blue means negative, and white is close to zero.  The lower right plot is colored according to the magnitude of $\vec{S}$, while the arrows show the direction of $\vec{S}$.  For $X_1 > 3$, $\vec{S}$ is so small that we probably cannot determine its direction reliably.  The Mach cone is shown in dashed green, and the solid blue line shows where the large distance analytic estimate of the diffusion wake profile reaches half its maximum value.}
\end{figure*}

Our numerical analysis was mostly done using $128^3$ grids for the three-dimensional FFT's of ${\cal E}^{\rm res}$ and $\vec{\cal S}^{\rm res}$, with wave-numbers $\vec{K}$ ranging from $-10$ to $10$ in each component.  The one-dimensional FFT's of the terms in ${\cal E}^{\rm IR}$ and $\vec{\cal S}^{\rm IR}$ with cubic denominators were performed on finer grids, usually with 1944 points, with $K_1$ ranging from $-20$ to $20$.  We chose $\mu_{\rm UV} = \mu_{\rm IR} = 1$.  Conservation was checked by comparing $\dot{\cal E}$ to $\nabla \cdot \vec{\cal S}$ in position space.  This is a stringent check because derivatives of numerically known functions tend to be noisy.  Conservation worked to within a few percent of $\dot{\cal E}$ except when $\dot{\cal E}$ became very small.  A summary of our numerical results for $v=0.75$ is shown in figure~\ref{BIGFIG}.  A more complete presentation of our numerical results is available \footnote{{\tt http://physics.princeton.edu/$\sim$ssgubser/papers\slash fourier/details.pdf}.}.

\section{Conclusions}

The Poynting vector describing the disturbances of a thermal plasma of ${\cal N}=4$ super-Yang-Mills theory from a moving quark exhibits a sonic boom and a diffusion wake, as expected on general grounds.  Using a natural split of the energy flow into sonic and diffusive modes, one finds---in the rest frame of the plasma---that energy is lost from the quark through sound modes and fed in toward the quark through the diffusion wake.  These two effects stand in the ratio $1+v^2 : -1$.  Thus the sonic boom is a strong effect, but the diffusion wake is also significant.

\begin{acknowledgments}
The work of S.~Gubser and S.~Pufu was supported in part by the Department of Energy under Grant No.\ DE-FG02-91ER40671, and by the Sloan Foundation.  A.~Yarom is supported in part by the German Science Foundation and by the Minerva foundation.
\end{acknowledgments}

\bibliographystyle{apsrev}
\bibliography{fourier}

\end{document}